\documentclass{article}

\usepackage{microtype}
\usepackage{graphicx}
\usepackage{subfigure}
\usepackage{booktabs}
\usepackage[inline]{enumitem}
\usepackage{hyperref}

\usepackage{amsmath}
\usepackage{amssymb}
\usepackage{mathtools}
\usepackage{amsthm}
\usepackage{bm}
\usepackage[capitalize,noabbrev]{cleveref}
\usepackage{siunitx}    
\usepackage{makecell}   
\usepackage{rotating}   
\usepackage{colortbl}   

\usepackage[accepted]{icml2025}
\theoremstyle{plain}

\theoremstyle{definition}

\theoremstyle{remark}

\usepackage[textsize=tiny]{todonotes}

\usepackage{xcolor}

\usepackage{multirow}
\usepackage{makecell} 

\usepackage{tikz}
\usetikzlibrary{shapes, arrows, positioning}
\usepackage[utf8]{inputenc}
\usepackage{pifont}

\icmltitlerunning{Agentic AI for Explainable Quantum Chemistry}

\begin{document}

\twocolumn[
\icmltitle{xChemAgents: Agentic AI for Explainable Quantum Chemistry}




\begin{icmlauthorlist}
\icmlauthor{Can Polat}{yyy}
\icmlauthor{Mehmet Tun\c{c}el}{sch1,sch2}
\icmlauthor{Mustafa Kurban}{sch}
\icmlauthor{Erchin Serpedin}{yyy}
\icmlauthor{Hasan Kurban}{comp}
\end{icmlauthorlist}

\icmlaffiliation{yyy}{Electrical \& Computer Engineering, Texas A\&M University, College Station, TX, USA}
\icmlaffiliation{comp}{College of Science and Engineering, Hamad Bin Khalifa University, Doha, Qatar}
\icmlaffiliation{sch}{Department of Prosthetics and Orthotics, Ankara University, Ankara, Turkey}
\icmlaffiliation{sch1}{Artificial Intelligence and Data Science Research Center, Istanbul Technical University, Istanbul, Türkiye}
\icmlaffiliation{sch2}{Department of Electrical \& Computer Engineering, Texas A\&M University at Qatar, Doha, Qatar}

\icmlcorrespondingauthor{Hasan Kurban}{hkurban@hbku.edu.qa}
\icmlcorrespondingauthor{Mustafa Kurban}{kurbanm@ankara.edu.tr}
\icmlkeywords{Machine Learning, ICML}

\vskip 0.3in
]



\printAffiliationsAndNotice{}  

\begin{abstract}
Recent progress in multimodal graph neural networks has demonstrated that augmenting atomic XYZ geometries with textual chemical descriptors can enhance predictive accuracy across a range of electronic and thermodynamic properties. However, naively appending large sets of heterogeneous descriptors often degrades performance on tasks sensitive to molecular shape or symmetry, and undermines interpretability. \textbf{xChemAgents} proposes a cooperative agent framework that injects physics-aware reasoning into multimodal property prediction. xChemAgents comprises two language-model-based agents: a \textit{Selector}, which adaptively identifies a sparse, weighted subset of descriptors relevant to each target, and provides a natural language rationale; and a \textit{Validator}, which enforces physical constraints such as unit consistency and scaling laws through iterative dialogue. On standard benchmark datasets, xChemAgents achieves up to a 22\% reduction in mean absolute error over the state-of-the-art baselines, while producing faithful, human-interpretable explanations. Experiment results highlight the potential of cooperative, self-verifying agents to enhance both accuracy and transparency in foundation-model-driven materials science. The implementation and accompanying dataset are available at \url{https://github.com/KurbanIntelligenceLab/xChemAgents}.
\end{abstract}

\section{Introduction}
\label{sec:intro}

First-principle electronic structure methods such as density functional theory (DFT) remain the gold standard for predicting bonding, defect energetics, and phase stability. However, the cubic wall-time scaling $\mathcal{O}(N^{3})$ with electron count due to repeated Hamiltonian diagonalization makes systematic studies of large or disordered systems impractical \cite{orio2009density, cohen2012challenges}. Density functional tight binding (DFTB) offers a more tractable alternative by tabulating two-center integrals and using minimal basis sets, but this reduction in cost comes at the expense of accuracy and transferability. Benchmark studies reveal large errors for long-range dispersion, many-body polarization, and strongly correlated states outside the fitted parameter domain \cite{hourahine2020dftb+, goyal2014molecular}. These trade-offs sustain the search for surrogate models that preserve quantum-mechanical fidelity while scaling to diverse, chemically rich systems.
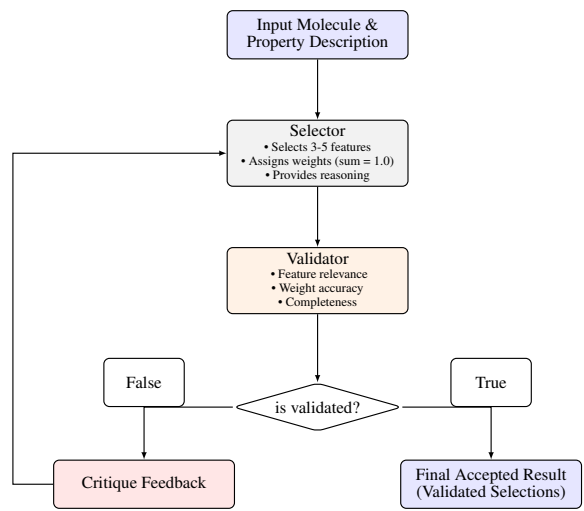
\begin{figure}
    \centering
    \resizebox{0.45\textwidth}{!}{
        \begin{tikzpicture}[
    node distance=1.5cm,
    every node/.style={draw, rounded corners, align=center, font=\footnotesize, minimum width=4.5cm, minimum height=1.2cm},
    every path/.style={draw, -latex, thick},
    decision/.style={diamond, aspect=3.5, minimum height=1.2cm},
    startend/.style={rectangle, rounded corners, fill=blue!10, minimum height=1.2cm},
    agent/.style={rectangle, fill=gray!10},
    critic/.style={rectangle, fill=orange!10}
]

\node[startend] (start) {\large{Input Molecule} \&\\\large{Property Description}};
\node[agent, below=of start] (agent) {\large{Selector}\\
• Selects 3-5 features\\
• Assigns weights (sum = 1.0)\\
• Provides reasoning};

\node[critic, below=of agent] (critic) {\large{Validator}\\\
• Feature relevance\\
• Weight accuracy\\
• Completeness};

\node[decision, below=of critic, yshift=-0.2cm, minimum width=3cm] (decision) {\large{is validated?}};

\node[startend, below right=1cm and 1cm of decision] (accepted) {\large{Final Accepted Result}\\\large{(Validated Selections)}};

\node[rectangle, below left=1cm and 1cm of decision, fill=red!10] (feedback) {\large{Critique Feedback}};

\draw (start) -- (agent);
\draw (agent) -- (critic);
\draw (critic) -- (decision);

\draw (decision) -| node[above, pos=0.50, minimum width=2cm] {\large{True}} (accepted);
\draw (decision) -| node[above,pos=0.50, minimum width=2cm] {\large{False}} (feedback);
\draw (feedback.west) -- ++(-1,0) |- (agent.west);
\end{tikzpicture}
    }
    \caption{Flowchart of the Selector–Validator pipeline for molecular feature selection. The process begins with an input molecule and its property description; the Selector agent identifies 3–5 candidate features, assigns normalized weights, and documents its reasoning. The Validator agent then evaluates feature relevance, weight accuracy, and overall completeness. If validation succeeds, the finalized, validated selections are output; if validation fails, critique feedback is returned to the Selector for iterative refinement.}
    \label{fig:multi_agent_validation}
\end{figure}

Graph neural networks (GNNs) have emerged as promising surrogates by operating directly on atomic graphs or periodic neighbor lists, achieving near-DFT accuracy with orders-of-magnitude speedups \cite{reiser2022graph}. Yet most state-of-the-art GNNs remain \textit{unimodal}, focusing solely on Cartesian geometry while ignoring rich textual metadata—empirical descriptors, spectral signatures, and systematic names—available in curated databases like PubChem \cite{pubchem2025}. Naïvely appending such descriptors can inflate the feature space, obscure physical relationships, and even degrade performance on symmetry-sensitive or long-range electrostatic tasks \cite{polat2025understanding}. These challenges exacerbate an ongoing interpretability deficit in materials machine learning, where node-level saliency maps rarely map cleanly to chemically meaningful insights \cite{oviedo2022interpretable}.

In parallel, recent advances in natural language processing have demonstrated that \textit{agentic AI}—systems composed of interacting large language model (LLM) agents—can solve complex reasoning tasks through cooperation, self-critique, and dialog-based refinement \cite{van2008multi, dorri2018multi, li2024survey}. Multi-agent orchestration enables autonomous scientific workflows for experiment planning, data analysis, and hypothesis generation, while reducing hallucination and improving robustness through inter-agent feedback loops \cite{xie2017multi}. These developments reframe feature selection for scientific surrogates not just as an optimization problem, but as an opportunity to embed domain knowledge and interpretability into the learning pipeline via deliberative, agentic reasoning.

This work presents \textbf{xChemAgents}, a cooperative multi-agent framework that embeds physics deliberation into multimodal molecular representation learning. A \emph{Selector} agent, implemented as a chemistry-tuned language model, queries a descriptor pool and proposes a compact, weighted subset of features for each target property, accompanied by a natural language rationale. A complementary \emph{Validator} agent enforces physical constraints—dimensional consistency, scaling laws, and chemical heuristics—and either approves the proposal or returns structured feedback for revision. This dialogue continues for up to three rounds, after which the validated features are fused with atomic embeddings and passed to a GNN surrogate. The result is a hybrid input that is not only more accurate, but also physically interpretable and auditable. The overall process is summarized in Figure~\ref{fig:multi_agent_validation}. xChemAgents exemplifies a shift toward agentic AI in scientific domains—transforming interpretability from a post-hoc explanation challenge into a multi-agent reasoning task grounded in domain principles.

The remainder of the paper is organized as follows. Section \ref{sec:bg} reviews background material. Section \ref{sec:method} describes the method. Section \ref{sec:experiments} presents experiments. Section \ref{sec:limits} discusses limitations. Section \ref{sec:discussion} concludes the paper.

\section{Background}
\label{sec:bg}

\subsection{Ab-Inito Simulations}
DFT maps the interacting many-electron problem onto a set of non-interacting single-particle equations solved self-consistently \cite{dft1}. Each self-consistency iteration requires diagonalizing the Kohn–Sham Hamiltonian—a “wall” that grows with the number of basis functions \cite{dft2}. While sparse-matrix techniques and linear-scaling ($\mathcal{O}(N)$) variants exploit locality to reduce formal complexity for large-gap (insulating) systems, metallic or highly disordered materials still demand dense algebra, so conventional DFT codes routinely incur wall-times of hours to days once system sizes exceed a few hundred atoms \cite{dft3}. Approximations such as DFTB perform a Taylor expansion of the total DFT energy around a reference density and tabulate two-centre integrals to achieve one- to two- orders of magnitude speed-ups over full DFT, but they inherit systematic errors—most notably missing long-range dispersion, incomplete many-body polarization, and challenges in treating strongly correlated states—without empirical corrections or higher-order expansions \cite{dft4}.

\subsection{Machine Learning for Materials Science}
Machine learning  surrogates have become the workhorse accelerators of contemporary materials modelling. Interatomic potentials and property predictors constructed with GNNs treat atoms as graph nodes and bonded or spatial neighbours as edges, enabling message-passing operations that capture complex many-body correlations directly from atomic coordinates. Models such as NequIP \cite{batzner20223}, TorchMD-Net \cite{tholke2022torchmd}, QuantumShellNet \cite{polat2025quantumshellnet}, and chemically focused CGNN variants \cite{cgnn} demonstrated that deep geometric learning can reproduce density-functional energies and forces at millisecond speed. Subsequent advances in equivariant design—exemplified by E(n)-GNN \cite{satorras2021n}, SE(3)-Transformer \cite{fuchs2020se}, PaiNN \cite{painn}, SphereNet \cite{spherenet}, and GotenNet \cite{gotennet}—embed rotational and permutational symmetry directly into the network, dramatically improving data efficiency and stability. More recently, multimodal encoders such as Pure2DopeNet \cite{polat2024multimodal} and CrysMMNet \cite{das2023crysmmnet} fuse structural graphs with textual and spectral channels, while generative pipelines like CDVAE \cite{xie2021crystal} and CrystalFlow \cite{luo2024crystalflow} learn to sample chemically valid molecules or crystals under explicit geometric constraints. Hierarchical pooling and attention mechanisms extend the effective receptive field to hundreds of atoms, yet—even with these innovations—the overwhelming majority of models remain geometry centric where they make little systematic use of the rich chemical metadata available in public repositories, and their latent features are difficult to map onto intuitive notions such as bond order, charge transfer or reaction centre. This opacity hampers error diagnosis, model trust and, ultimately, adoption by domain scientists.

\subsection{Multi-Agent Systems}
Multi-agent systems provide a principled framework for modularizing complex scientific tasks \cite{michel2018multi}. Recent work has demonstrated the integration of agents with domain-specific tools—such as one agent orchestrating simulations, another tracking convergence criteria, and a third triaging results and suggesting follow-up experiments \cite{song2025multiagent}. Cooperative protocols have been successfully applied in diverse scientific domains, including autonomous synthesis planning, closed-loop catalyst discovery, and the inverse design of battery electrolytes \cite{jia2025closed}.

Within representation learning, a prominent Selector–Validator pattern has emerged: a reasoning agent proposes candidate features or hypotheses, while a critic agent validates them against physical constraints such as dimensional consistency or known scaling laws. The xChemAgents framework exemplifies this paradigm by transforming descriptor selection from a static preprocessing step into an iterative, agent-driven dialogue. This leads to sparse, physically grounded, and interpretable representations that serve as inputs to downstream GNN-based surrogate models \cite{zhang2025mars}.

\section{Method}
\label{sec:method}
\subsection{Agentic Descriptor Selection \& Validation}
\label{sec:agentic}

Let $\mathcal{D} = \{\phi_{1}, \dots, \phi_{m}\}$ denote the full bank of textual descriptors provided with the enriched QM9 dataset \cite{ruddigkeit2012enumeration,ramakrishnan2014quantum}, where $m = 9$ in this study. Each descriptor $\phi_{k}$ is embedded once using a frozen CLIP encoder \cite{clip}, yielding a fixed vector representation $\bm{\phi}_{k} \in \mathbb{R}^{d_\phi}$.

\textbf{Selector Agent.} Given an input molecule $x$ and a target property $y$, the Selector agent is a function  
\[
\mathcal{A}_{\mathrm{sel}} : (x, y) \mapsto (S, \mathbf{w}, r_{\mathrm{sel}}),
\]  
which outputs:
\begin{enumerate}[label=(\arabic*)]
  \item a rationale $r_{\mathrm{sel}}$ in natural language (chain-of-thought),
  \item a selected subset $S = \{\phi_{k_1}, \dots, \phi_{k_p}\} \subset \mathcal{D}$ with $p \in \{3, 4, 5\}$,
  \item a weight vector $\mathbf{w} \in \Delta^{p-1}$, where $\Delta^{p-1} = \left\{ \mathbf{w} \in \mathbb{R}^p : \sum_{i=1}^p w_i = 1, \, w_i \ge 0 \right\}$ is the $(p{-}1)$-dimensional probability simplex.
\end{enumerate}
\textbf{Validator Agent.} The Validator agent  
\[
\mathcal{A}_{\mathrm{val}} : (S, \mathbf{w}, y) \mapsto (v, r_{\mathrm{val}})
\]  
accepts the selected descriptors $S$, their weights $\mathbf{w}$, and the target property $y$, returning:
\begin{itemize}
  \item a binary verdict $v \in \{0, 1\}$ indicating acceptance ($1$) or rejection ($0$),
  \item a critique $r_{\mathrm{val}}$ in natural language explaining the reasoning behind $v$.
\end{itemize}

The Validator is guided by structured prompts encoding:
\begin{itemize}
  \item[(i)] unit consistency between each $\phi_k$ and $y$,
  \item[(ii)] adherence to known scaling relations (e.g., Koopmans’ theorem for HOMO/LUMO predictions),
  \item[(iii)] sparsity and complementarity heuristics to prevent redundancy and encourage diversity among descriptors.
\end{itemize}

If $v = 0$, the critique $r_{\mathrm{val}}$ is appended to the dialogue history, and the Selector re-samples. This dialogue loop runs for at most $L = 3$ rounds (see Algorithm~\ref{alg:xChemAgents}).

\begin{algorithm}[tb]
   \caption{xChemAgents: End-to-End Physics-Vetted Inference}
   \label{alg:xChemAgents}
\begin{algorithmic}[1]
   \STATE \textbf{Input:} atomic positions $\{\mathbf r_i\}$, atomic numbers $\{Z_i\}$, target property $y$
   \STATE \textbf{Require:} descriptor bank $\mathcal D$, selector agent $\mathcal A_{\text{sel}}$, validator $\mathcal A_{\text{val}}$, equivariant GNN $\mathcal G_\theta$, fusion head $\mathcal F_\theta$
   \STATE \textbf{Output:} prediction $\hat y$, accepted feature set $\hat S$, weights $\hat{\mathbf w}$, rationales

      \FOR{$\ell=1$ \textbf{to} $L$}
        \STATE $(S,\mathbf w,r_{\text{sel}})\gets \mathcal A_{\text{sel}}(y,\mathcal D)$
        \STATE $(v,r_{\text{val}})\gets \mathcal A_{\text{val}}(S,\mathbf w,y)$
        \IF{$v$}
          \STATE $\hat S\gets S$, $\hat{\mathbf w}\gets\mathbf w$
          \STATE \textbf{break}
        \ENDIF
      \ENDFOR
   \STATE $\mathbf t_{\text{phys}} \leftarrow \sum_{\phi_k \in \hat S} \hat w_k \,\text{CLIP}(\phi_k)$

   \STATE initialise $\mathbf h_i^{(0)} \leftarrow \text{onehot}(Z_i)$
   \FOR{$t = 0$ \textbf{to} $T-1$}
      \FORALL{edges $(i,j)$ with $d_{ij} < r_c$}
         \STATE $\mathbf m_{ij}^{(t)} \leftarrow \text{TP}\!\bigl(\mathbf h_i^{(t)},\mathbf h_j^{(t)},\text{RBF}(d_{ij})\bigr)$
      \ENDFOR
      \STATE $\mathbf h_i^{(t+1)} \leftarrow \mathbf h_i^{(t)} + \sum_{j\in\mathcal N(i)} \mathbf m_{ij}^{(t)}$
   \ENDFOR
   \STATE $\mathbf g \leftarrow \sum_i \sigma\!\bigl(W_{\text{gate}}\mathbf h_i^{(T)}\bigr)\,\mathbf h_i^{(T)}$

   \STATE $\tilde{\mathbf t} \leftarrow W_t P \mathbf t_{\text{phys}}$;\quad $\tilde{\mathbf g} \leftarrow W_g \mathbf g$
   \STATE normalise $(\tilde{\mathbf g},\tilde{\mathbf t})$
   \STATE $\mathbf z \leftarrow \sigma\!\bigl(W[\tilde{\mathbf g}\!\|\!\tilde{\mathbf t}] + b\bigr)$
   \STATE $\mathbf f \leftarrow \mathbf z \odot \tilde{\mathbf g} + (1-\mathbf z) \odot \tilde{\mathbf t}$
   \STATE $\hat y \leftarrow \mathcal F_\theta(\mathbf f)$
\end{algorithmic}
\end{algorithm}
\textbf{Physics-aware texual embedding.} Given the accepted descriptor set $\hat{S} = \{\phi_{k_1}, \dots, \phi_{k_p}\}$ and corresponding normalized weights $\hat{w}_j$, the final physics-aware text embedding is computed as a weighted average:
\begin{equation}
\mathbf{t}_{\mathrm{phys}} = \sum_{j=1}^{p} \hat{w}_{j}\, \bm{\phi}_{k_j},
\label{eq:tphys}
\end{equation}
where $\bm{\phi}_{k_j} \in \mathbb{R}^{d_\phi}$ denotes the frozen CLIP embedding of descriptor $\phi_{k_j}$. This representation replaces the original CLIP text vector in the multimodal fusion module (Section~\ref{sec:fusion}), ensuring that only physics-vetted information influences the downstream prediction.

\subsection{Equivariant Geometric Encoder}
\label{sec:gnn}

Each atom is represented as a node $i$ with an initial feature vector $\mathbf{h}_{i}^{(0)} = \mathrm{onehot}(Z_{i})$, where $Z_{i}$ is the atomic number. Edges $(i, j)$ connect all atomic pairs within a cutoff radius $r_{c}$, based on interatomic distances $d_{ij} = \|\mathbf{r}_i - \mathbf{r}_j\|$.

Distances are expanded using Bessel radial basis functions (RBFs):
\[
\bm{\rho}(d_{ij}) = [\rho_1(d_{ij}), \dots, \rho_B(d_{ij})],
\]
where $\rho_b(\cdot)$ denotes the $b$-th basis function. Angular features $\theta_{ijk}$ are similarly embedded using real spherical harmonics $Y_{\ell m}(\theta_{ijk})$ to capture 3-body geometric interactions.

At message-passing step $t \in \{0, \dots, T{-}1\}$, node embeddings are updated as:
\begin{align}
\mathbf{m}_{ij}^{(t)} &= \mathrm{TP}(\mathbf{h}_{i}^{(t)}, \mathbf{h}_{j}^{(t)}, \bm{\rho}(d_{ij})), \\
\mathbf{h}_{i}^{(t+1)} &= \mathbf{h}_{i}^{(t)} + \sum_{j \in \mathcal{N}(i)} \mathbf{m}_{ij}^{(t)},
\end{align}
where $\mathrm{TP}$ denotes a learnable tensor product that mixes irreducible representations of $\mathrm{SO}(3)$, ensuring equivariance under global rotations.

After $T$ message-passing layers, the final node embeddings are aggregated via gated summation:
\begin{equation}
\mathbf{g} = \sum_{i} \sigma(W_{\mathrm{gate}} \mathbf{h}_{i}^{(T)}) \odot \mathbf{h}_{i}^{(T)} \in \mathbb{R}^{n},
\label{eq:geomvec}
\end{equation}
where $W_{\mathrm{gate}} \in \mathbb{R}^{n \times d}$ is a learnable projection matrix, $\sigma(\cdot)$ denotes the sigmoid activation, and $\odot$ indicates elementwise multiplication.

\subsection{Multimodal Fusion and Prediction}
\label{sec:fusion}

The text-derived embedding $\mathbf{t}_{\mathrm{phys}}$ (see Eq.~\ref{eq:tphys}) is projected into a $d$-dimensional latent space:
\begin{equation}
\mathbf{t}_{p} = P\,\mathbf{t}_{\mathrm{phys}}, \quad P \in \mathbb{R}^{d \times 768}.
\end{equation}
The geometric embedding $\mathbf{g}$ (from Eq.~\ref{eq:geomvec}) is projected using learned linear transformations:
\begin{equation}
\mathbf{g}_{\mathrm{proj}} = W_{g} \mathbf{g}, \quad
\mathbf{t}_{\mathrm{proj}} = W_{t} \mathbf{t}_{p},
\end{equation}
where $W_g, W_t \in \mathbb{R}^{d \times d}$. Both embeddings are normalized using layer normalization:
\[
\tilde{\mathbf{g}} = \mathrm{LN}(\mathbf{g}_{\mathrm{proj}}), \quad 
\tilde{\mathbf{t}} = \mathrm{LN}(\mathbf{t}_{\mathrm{proj}}).
\]

A gating mechanism computes elementwise fusion weights:
\[
\mathbf{z} = \sigma\left(W[\tilde{\mathbf{g}} \| \tilde{\mathbf{t}}] + b\right),
\]
where $W \in \mathbb{R}^{d \times 2d}$ is a learned matrix, $b \in \mathbb{R}^{d}$ is a bias vector, $\|$ denotes concatenation, and $\sigma(\cdot)$ is the sigmoid function.

The fused representation is computed as:
\begin{equation}
\mathbf{f} = \mathbf{z} \odot \tilde{\mathbf{g}} + (1 - \mathbf{z}) \odot \tilde{\mathbf{t}},
\end{equation}
where $\odot$ denotes elementwise multiplication. This vector $\mathbf{f} \in \mathbb{R}^d$ is passed to a multi-layer perceptron $f_{\theta}$ to predict the target property:
\begin{equation}
\hat{y} = f_{\theta}(\mathbf{f}).
\end{equation}

The model is trained to minimize the mean squared error between $\hat{y}$ and the ground-truth property. During training and inference, both agents remain frozen. The agent dialogue process (Algorithm~\ref{alg:xChemAgents}) introduces an average overhead of approximately 0.2 seconds per sample but reduces the descriptor channel from 9 to a maximum of 5. This enables improved interpretability and physical alignment without compromising predictive throughput.

\section{Experiments}
\label{sec:experiments}

\textbf{Dataset.} The QM9 benchmark consists of approximately 134k small organic molecules stored as XYZ files, which capture only their three-dimensional geometries. To provide the chemical context absent from atomic coordinates, each molecule is augmented with metadata retrieved from PubChem, including IUPAC names, molecular formulas and masses, XLogP, counts of hydrogen-bond donors and acceptors, rotatable bond counts, topological polar surface area, formal charge, synonyms, and selected spectral annotations. This results in a multimodal dataset that integrates geometric and textual descriptors to enable richer property prediction. Molecules lacking any of the required PubChem fields are excluded from the final corpus.

\textbf{Implementations.} Evaluation is conducted on the QM9 benchmark using four backbone architectures that span diverse regions of the contemporary GNN design space: (i) \textbf{SchNet}; (ii) \textbf{DimeNet++}, which employs directional message passing and spherical harmonics; (iii) \textbf{Equiformer}, an E(3)-equivariant graph-attention transformer; and (iv) \textbf{FAENet}, a frame-averaging network without explicit \(\mathrm{SO}(3)\) constraints. To ensure architectural consistency, all backbones are equipped with a shared gated-fusion block, isolating the impact of the agentic feature selector from variations in model capacity. Networks are implemented using PyTorch \cite{pytorch}, PyTorch Geometric \cite{torch_geo}, and the official repositories. Optimization is performed using the ADAM optimizer \cite{adam} with an \(\ell_1\) loss, a learning rate of \(10^{-3}\), a batch size of 64, and 35 training epochs. Performance is reported as the mean over three randomized folds (\(k=3\)) to account for dataset variability. The agentic system is implemented using LangChain, with Ollama serving as the backbone model and CLIP as the tokenizer, integrated via the Transformers library \cite{wolf2019huggingface}.

\textbf{Gated-fusion layer.}  
Each backbone outputs a graph-level embedding \(\mathbf{g} \in \mathbb{R}^{128}\), which is fused with a physics-vetted CLIP embedding \(\mathbf{t}_{\text{phys}} \in \mathbb{R}^{32}\). The fusion is performed via a learnable gating mechanism:
\[
\mathbf{f} = \sigma\left(W\left[\mathbf{g} \| \mathbf{t}_{\text{phys}}\right]\right) \odot W_g \mathbf{g}
+ \left(1 - \sigma(\cdot)\right) \odot W_t \mathbf{t}_{\text{phys}},
\]
where \([\cdot \| \cdot]\) denotes concatenation, \(\sigma(\cdot)\) is the sigmoid function, and \(\odot\) denotes element-wise multiplication. The resulting multimodal vector \(\mathbf{f}\) is passed to a two-layer MLP for final property prediction.

\textbf{SchNet.}  
The SchNet configuration uses six interaction blocks, a hidden size of 128, 50 Gaussian filters, and a cutoff radius of \(10\,\text{\AA}\). Global add-pooling is applied to produce a 128-dimensional graph-level embedding, which is fused with the CLIP vector via the gated-fusion layer. A single linear layer outputs the scalar prediction.

\textbf{DimeNet++.}  
DimeNet++ is configured with hyperparameters scaled down by half relative to the original implementation: three interaction blocks, 128 hidden channels, a spherical rank of \(l_{\max} = 7\), radial basis size of 6, and a cutoff radius of \(5\,\text{\AA}\). The default readout is replaced with the gated-fusion layer followed by a linear prediction head.

\textbf{Equiformer.}  
The Equiformer employs four attention layers with two heads and uses 50\% fewer representation channels than the original model, resulting in approximately 4.8M parameters—comparable to FAENet. The graph-level embedding has a dimensionality of 256, which is fused with a 32-dimensional CLIP branch and passed through a linear layer for prediction.

\textbf{FAENet.}  
FAENet is configured with a cutoff radius of \(6\,\text{\AA}\), two interaction blocks, a hidden dimension of 128, and 20 Gaussian distance channels. The textual CLIP embedding (768-dimensional) is reduced to 32 dimensions via a two-layer projection head before being fused with the graph representation.

\begin{figure}
    \centering
    \includegraphics[width=0.95\linewidth]{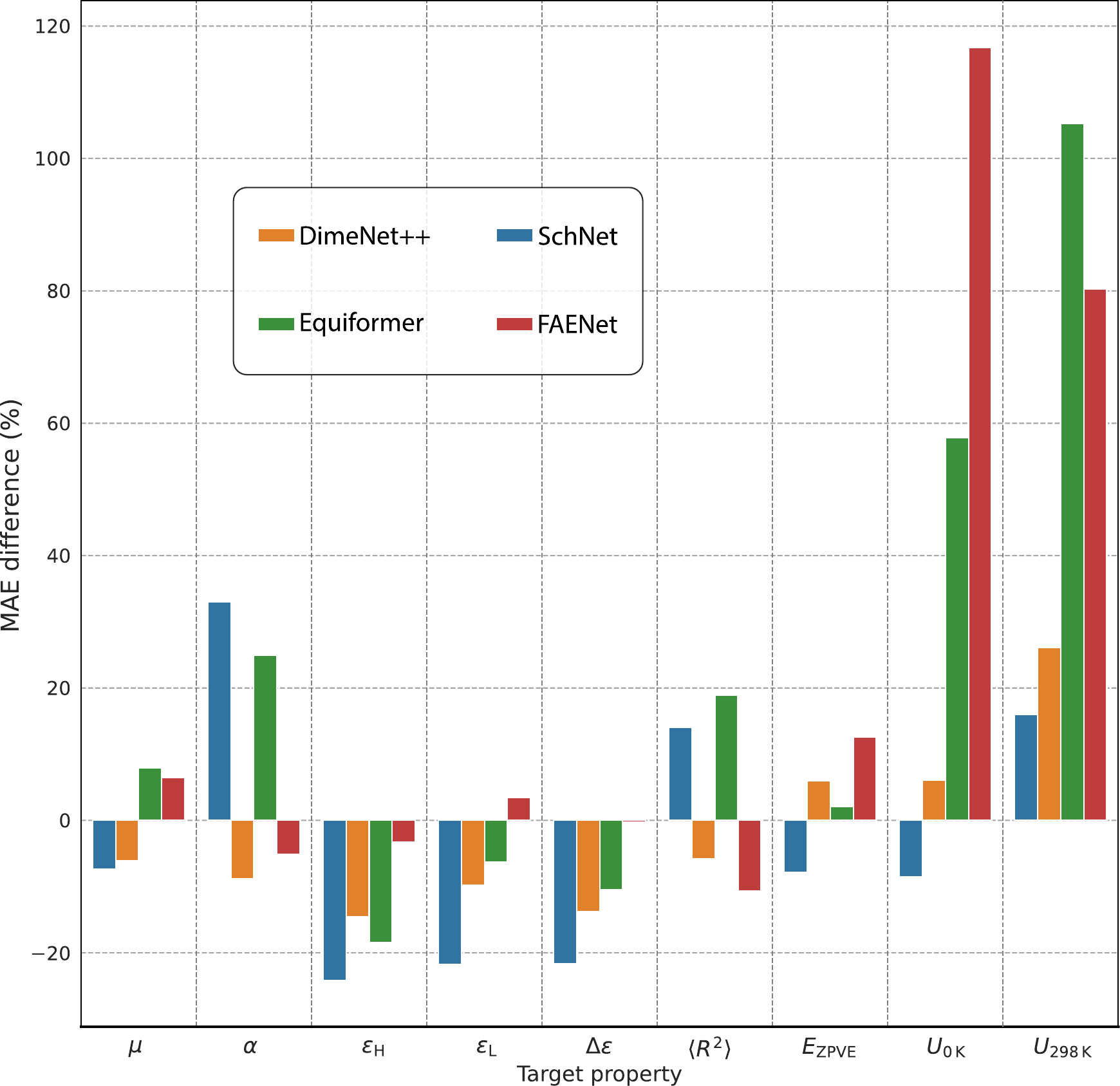}
    \caption{Percentile change in MAE across models for each target property. Positive values denote an increase in error relative to the baseline, while negative values indicate an improvement in the model’s performance.}
    \label{fig:percent_changes}
\end{figure}

\begin{table*}[ht]
\centering
\caption{Mean Absolute Error (MAE; lower is better) for molecular property prediction across baseline architectures and their xChemAgents-enhanced variants. Models augmented with xChemAgents are highlighted in \textbf{bold}, reflecting consistent performance improvements through physics-aware cooperative reasoning.}
\label{tbl:maeValues}
\footnotesize
\setlength{\tabcolsep}{6pt}
\renewcommand{\arraystretch}{1.15}
\begin{tabular}{@{}lcccccccc@{}}
\toprule
\textbf{Target} & 
\multicolumn{2}{c}{SchNet} & 
\multicolumn{2}{c}{DimeNet++} & 
\multicolumn{2}{c}{Equiformer} & 
\multicolumn{2}{c}{FAENet} \\
& Base & xChemAgents & Base & xChemAgents & Base & xChemAgents & Base & xChemAgents \\
\midrule
$\mu$ (D)                          & 0.2308 & \textbf{0.2139} & 0.1747 & \textbf{0.1640} & 0.1154 & 0.1245 & 0.2650 & 0.2820 \\
$\alpha$ ($\text{\AA}^3$)         & 0.3154 & 0.4194 & 0.3762 & \textbf{0.3431} & 0.2887 & 0.3606 & 0.4925 & \textbf{0.4675} \\
$\varepsilon_{\text{H}}$ (eV)     & 0.1790 & \textbf{0.1358} & 0.1317 &\textbf{0.1125} & 0.1380 & \textbf{0.1126} & 0.1424 & \textbf{0.1377} \\
$\varepsilon_{\text{L}}$ (eV)     & 0.1312 & \textbf{0.1027} & 0.0891 & \textbf{0.0804} & 0.0814 & \textbf{0.0763} & 0.0980 & 0.1014 \\
$\Delta\varepsilon$ (eV)          & 0.2271 & \textbf{0.1779} & 0.1726 & \textbf{0.1489} & 0.1627 & \textbf{0.1457} & 0.1932 & \textbf{0.1928} \\
$\langle R^2 \rangle$ (bohr$^2$)  & 5.4741 & 6.2421 & 9.6462 & \textbf{9.0899} & 5.5983 & 6.6559 & 15.0068 & \textbf{13.4011} \\
$E_{\text{ZPVE}}$ (eV)            & 0.0115 & \textbf{0.0106} & 0.0067 & 0.0071 & 0.0094 & 0.0096 & 0.0095 & 0.0107 \\
$U_{0\,\mathrm{K}}$ (eV)          & 30.9582 & \textbf{28.3283} & 39.7068 & 42.1025 & 12.3913 & 19.5572 & 9.8065 & 21.2594 \\
$U_{298\,\mathrm{K}}$ (eV)        & 28.1507 & 32.6541 & 32.7486 & 41.3099 & 13.0388 & 26.7677 & 11.0204 & 19.8619 \\
\bottomrule
\end{tabular}
\end{table*}

\begin{table*}[t]
\centering
\caption{Summary of descriptor relevance across target properties. Row 1 reports the number of times each descriptor was selected by the Selector agent across runs. Row 2 shows the average normalized importance weights assigned to each descriptor. Best values in each row are shown in \textbf{bold}, and second-best are \underline{underlined}.}
\label{tab:feature_selection_summary}
\footnotesize
\setlength{\tabcolsep}{4pt}
\renewcommand{\arraystretch}{1.15}
\resizebox{\linewidth}{!}{
\begin{tabular}{@{}llccccccccc@{}}
\toprule
\textbf{Target} & \textbf{Metric} & IUPAC & Formula & Molecular Weight & XLogP & H-Bond Donors & H-Bond Acceptors & Rotatable Bonds & PSA & Synonyms \\
\midrule
\multirow{2}{*}{$\mu$} 
& Selection Count & 2340 & 2734 & \textbf{21597} & 13378 & 14935 & 16893 & 8134 & \underline{18146} & 79 \\
& Normalized Importance & 0.26 & 0.26 & \underline{0.28} & 0.24 & 0.22 & 0.24 & 0.23 & \textbf{0.29} & 0.18 \\
\midrule
\multirow{2}{*}{$\alpha$} 
& Selection Count & 951 & 2328 & \textbf{22223} & 14576 & 6458 & 9241 & \underline{18344} & 15816 & 81 \\
& Normalized Importance & 0.26 & 0.27 & \textbf{0.31} & 0.26 & 0.20 & 0.21 & 0.25 & \underline{0.28} & 0.18 \\
\midrule
\multirow{2}{*}{$\varepsilon_{\text{H}}$} 
& Selection Count & 2661 & 3548 & \textbf{21205} & 15724 & 9942 & 10478 & \underline{16430} & 11820 & 190 \\
& Normalized Importance & 0.27 & 0.27 & \textbf{0.34} & \underline{0.31} & 0.23 & 0.23 & 0.30 & 0.27 & 0.21 \\
\midrule
\multirow{2}{*}{$\varepsilon_{\text{L}}$} 
& Selection Count & 2185 & 3194 & \textbf{19068} & 16700 & 9680 & 13294 & \underline{17484} & 10650 & 175 \\
& Normalized Importance & 0.26 & 0.26 & \underline{0.28} & \textbf{0.82} & 0.22 & 0.23 & 0.25 & 0.26 & 0.20 \\
\midrule
\multirow{2}{*}{$\Delta\varepsilon$} 
& Selection Count & 1455 & 2537 & \textbf{20497} & 16425 & 8570 & 9592 & \underline{17726} & 10453 & 157 \\
& Normalized Importance & 0.26 & \textbf{0.64} & \underline{0.31} & 0.26 & 0.23 & 0.22 & 0.28 & 0.28 & 0.18 \\
\midrule
\multirow{2}{*}{$\langle R^2 \rangle$} 
& Selection Count & 1001 & 1544 & \textbf{22064} & 7553 & 3022 & 5992 & 21634 & \underline{21964} & 117 \\
& Normalized Importance & 0.27 & 0.28 & \textbf{0.39} & 0.25 & 0.21 & 0.22 & 0.28 & \underline{0.29} & 0.19 \\
\midrule
\multirow{2}{*}{$E_{\text{ZPVE}}$} 
& Selection Count & 1115 & 2594 & \textbf{23092} & 9477 & 10249 & 10335 & \underline{16997} & 15379 & 113 \\
& Normalized Importance & 0.26 & 0.28 & \textbf{0.35} & 0.24 & 0.27 & 0.25 & 0.26 & \underline{0.31} & 0.18 \\
\midrule
\multirow{2}{*}{$U_{0\,\mathrm{K}}$} 
& Selection Count & 1116 & 2668 & \textbf{22606} & 11769 & 10563 & 10494 & \underline{15230} & 11980 & 83 \\
& Normalized Importance & 0.26 & \textbf{0.44} & \underline{0.35} & 0.29 & 0.23 & 0.23 & 0.28 & 0.27 & 0.17 \\
\midrule
\multirow{2}{*}{$U_{298\,\mathrm{K}}$} 
& Selection Count & 728 & 1614 & \textbf{23925} & 14419 & 12016 & 9394 & \underline{15722} & 13073 & 74 \\
& Normalized Importance & 0.25 & \underline{0.28} & \textbf{0.33} & 0.26 & 0.22 & 0.23 & 0.24 & 0.27 & 0.18 \\
\bottomrule
\end{tabular}}
\end{table*}

\subsection{Effect on Loss}
The inclusion of textual annotations—including IUPAC names, chemical formulas, molecular weights, rotatable bond counts, polar surface areas, hydrogen-bond donor/acceptor counts, and synonyms—introduces high-level information related to functional groups and molecular polarity. These descriptors correlate strongly with frontier orbital behavior, resulting in consistent reductions in mean absolute error (MAE) across all backbones for HOMO energy \(\varepsilon_{\text{H}}\), LUMO energy \(\varepsilon_{\text{L}}\), and the HOMO–LUMO gap \(\Delta\varepsilon\), as shown in Table~\ref{tbl:maeValues}. Smaller but noticeable improvements are also observed for the dipole moment \(\mu\), whose magnitude reflects charge separation patterns captured by the same descriptors.

In contrast, internal energies \(U_{0\,\mathrm{K}}\) and \(U_{298\,\mathrm{K}}\), which depend on full potential-energy surfaces and low-frequency vibrational modes, show limited benefit. Here, the high-level descriptors contribute little and often act as correlated noise, resulting in stagnating or increased MAE—most visibly for DimeNet++, Equiformer, and FAENet.

The location and frequency of fusion modulate these trends. SchNet and DimeNet++ apply the annotation vector only at the global readout stage, allowing auxiliary information to influence the output without disrupting symmetry-aware message passing. This late-fusion strategy leads to net improvements on five of nine targets. Equiformer, by contrast, injects the annotation vector at every attention layer, repeatedly mixing non-geometric tokens into equivariant updates. This early-fusion approach benefits orbital-related tasks but degrades geometry-dominated properties such as isotropic polarizability \(\alpha\) and electronic spatial extent \(\langle R^{2} \rangle\). FAENet adopts an intermediate strategy and exhibits mixed behavior—improving predictions for HOMO energies while reducing accuracy for dipole moments and LUMO levels.

These results suggest that simple text-based descriptors are most effective for properties governed by electronic distribution, whereas thermodynamic quantities demand richer, physics-aware encodings such as force constants or vibrational fingerprints. The findings also underscore the importance of aligning the abstraction level of auxiliary information with the backbone’s inductive biases: the same cue that improves one prediction can degrade another if integrated incongruently. A percentile-based comparison is provided in Figure~\ref{fig:percent_changes}.

\subsection{xChemAgents Decisions}

Table~\ref{tab:feature_selection_summary} summarizes the behavior of the \textit{Selector} agent across all tasks. The first row records the frequency with which each textual descriptor is selected, while the second row reports the mean normalized weight assigned when the feature is included.

A property-agnostic trend emerges: \textbf{molecular weight} overwhelmingly dominates the selection statistics, being chosen over 20,000 times across all electronic and thermodynamic targets—far more than any other descriptor. This reflects the agent’s reliance on a coarse proxy for molecular size and electron count in the absence of more specific physical signals.

However, the normalized weights reveal a more nuanced behavior. High selection frequency does not necessarily imply high influence. For several tasks—most notably LUMO energy \(\varepsilon_{\text{L}}\) and the HOMO–LUMO gap \(\Delta\varepsilon\)—less frequently chosen descriptors are assigned significantly larger weights when selected. This suggests that the agent defaults to molecular weight for general coverage but relies on more specialized descriptors to refine predictions where they matter most.

\textbf{Electronic descriptors.}
For LUMO energy \(\varepsilon_{\text{L}}\), the lipophilicity index XLogP receives the highest average weight (0.82), nearly three times greater than any other descriptor, despite being selected only 16,700 times compared to 19,068 selections for molecular weight. A high XLogP typically indicates the presence of extended hydrophobic and often conjugated fragments, which stabilize low-lying unoccupied orbitals—explaining its elevated influence on \(\varepsilon_{\text{L}}\).

For the HOMO–LUMO gap \(\Delta\varepsilon\), a complementary trend emerges: the \textbf{gross formula} achieves the highest mean weight (0.64), suggesting that elemental composition—especially the inclusion of heteroatoms—serves as a dominant discriminator once molecular size is accounted for.

Properties such as dipole moment \(\mu\) and isotropic polarizability \(\alpha\) depend on broader polarity cues. Here, polar surface area and hydrogen-bond donor/acceptor counts are assigned moderate but consistent weights, reflecting their connection to charge separation and electron cloud deformability. In contrast, \texttt{Synonyms}, being purely linguistic metadata, are rarely selected and consistently assigned the lowest weights, underscoring their limited physical relevance.

\textbf{Thermodynamic quantities.}
Zero-point vibrational energy \(E_{\text{ZPVE}}\) and internal energies \(U_{0\,\mathrm{K}}\) and \(U_{298\,\mathrm{K}}\) exhibit a strong selection bias toward molecular weight, consistent with the scaling of vibrational degrees of freedom with system size. However, their average assigned weights (all \(\le 0.35\)) remain notably lower than those observed for frontier-orbital tasks, reinforcing the earlier observation that high-level string descriptors offer limited signal for properties governed by detailed potential-energy surfaces.

For these targets, the \textit{Selector} agent distributes weights more evenly across descriptors such as XLogP, hydrogen-bond statistics, and polar surface area, indicating that no single textual feature provides dominant predictive power. This behavior suggests that the agent is effectively “hedging” by forming diluted ensembles—a strategy consistent with the modest or negative MAE changes observed in Table~\ref{tbl:maeValues}.

Overall, the feature selection patterns support the conclusion that simple text-based descriptors are most effective for electronic observables. In contrast, thermodynamic properties appear to require richer, physics-aware representations—such as force constants or vibrational fingerprints—to achieve comparable predictive improvements.

\section{Limitations}
\label{sec:limits}
xChemAgents is currently evaluated only within the constrained chemical space of QM9, which limits generalization to larger, charged, or inorganic species. The system also relies on the availability and quality of PubChem descriptors; in cases where these metadata are missing, predictions revert to geometry-only inputs.

The Validator agent enforces primarily first-order physical constraints, such as unit consistency and selected scaling laws. More nuanced issues—such as quantum interference effects or stereochemical errors—are not explicitly addressed.

Additionally, the three-round Selector–Validator dialogue introduces modest computational overhead, adding latency per molecule. This study focuses exclusively on ground-state scalar properties; extensions to vector-valued outputs, forces, excited-state observables, and kinetic parameters are left for future work.

\section{Discussion \& Conclusion}
\label{sec:discussion}

Augmenting geometry-based backbones with a compact set of string-level descriptors consistently improved predictions for electronically governed properties—such as dipole moment, frontier orbital energies, and the HOMO–LUMO gap—while leaving thermodynamic quantities largely unaffected or slightly degraded. Late-fusion architectures like SchNet and DimeNet++ integrated these auxiliary signals most effectively, indicating that global chemical context is best introduced after symmetry-aware message passing. In contrast, early fusion disrupted geometric invariances and led to diminished accuracy on shape-sensitive targets.

Feature selection patterns reinforce these observations. Although molecular weight was the most frequently selected descriptor, the highest average weights shifted toward chemically informative features—XLogP for LUMO energy and elemental formula for the HOMO–LUMO gap—highlighting the importance of polarity and composition in modulating electronic structure. For thermodynamic targets, no single textual feature exhibited dominant influence, emphasizing the need for richer, physics-grounded descriptors (e.g., force constants, vibrational modes) to achieve similar improvements.

In summary, this study demonstrates the potential of cooperative, agentic AI to selectively extract and integrate multimodal chemical knowledge. xChemAgents offers a step toward interpretable, physics-aware surrogates that can be extended beyond ground-state scalars to richer tasks in materials and molecular modeling.

\section*{Software and Data}

The xChemAgents toolkit—including source code, curated textual descriptors, and the complete Selector–Validator implementation—is released under the MIT license and is available at the following link: \url{https://github.com/KurbanIntelligenceLab/xChemAgents}. All datasets and annotations are derived exclusively from open-access resources and are free of proprietary restrictions.

\section*{Impact Statement}

xChemAgents integrates physics-aware language agents with GNNs to enable faster and more interpretable virtual screening, offering potential benefits for green catalyst design and drug discovery. By reducing descriptor overhead and producing transparent, logged rationales, the framework supports reproducibility and regulatory compliance. Key risks—including biased descriptor pools, overreliance on automated explanations, and the carbon footprint of training—can be mitigated through the use of balanced datasets, periodic rule audits, and energy-efficient fine-tuning strategies.

\bibliography{example_paper}

\begin{thebibliography}{40}
\providecommand{\natexlab}[1]{#1}
\providecommand{\url}[1]{\texttt{#1}}
\expandafter\ifx\csname urlstyle\endcsname\relax
  \providecommand{\doi}[1]{doi: #1}\else
  \providecommand{\doi}{doi: \begingroup \urlstyle{rm}\Url}\fi

\bibitem[Aykent \& Xia(2025)Aykent and Xia]{gotennet}
Aykent, S. and Xia, T.
\newblock {GotenNet: Rethinking Efficient 3D Equivariant Graph Neural Networks}.
\newblock In \emph{The Thirteenth International Conference on LearningRepresentations}, 2025.
\newblock URL \url{https://openreview.net/forum?id=5wxCQDtbMo}.

\bibitem[Batzner et~al.(2022)Batzner, Musaelian, Sun, Geiger, Mailoa, Kornbluth, Molinari, Smidt, and Kozinsky]{batzner20223}
Batzner, S., Musaelian, A., Sun, L., Geiger, M., Mailoa, J.~P., Kornbluth, M., Molinari, N., Smidt, T.~E., and Kozinsky, B.
\newblock E (3)-equivariant graph neural networks for data-efficient and accurate interatomic potentials.
\newblock \emph{Nature communications}, 13\penalty0 (1):\penalty0 2453, 2022.

\bibitem[Cheng et~al.(2021)Cheng, Zhang, and Dong]{cgnn}
Cheng, J., Zhang, C., and Dong, L.
\newblock A geometric-information-enhanced crystal graph network for predicting properties of materials.
\newblock \emph{Communications Materials}, 2\penalty0 (1):\penalty0 92, 2021.

\bibitem[Cohen et~al.(2012)Cohen, Mori-S{\'a}nchez, and Yang]{cohen2012challenges}
Cohen, A.~J., Mori-S{\'a}nchez, P., and Yang, W.
\newblock Challenges for density functional theory.
\newblock \emph{Chemical Reviews}, 112\penalty0 (1):\penalty0 289--320, 2012.

\bibitem[Coors et~al.(2018)Coors, Condurache, and Geiger]{spherenet}
Coors, B., Condurache, A.~P., and Geiger, A.
\newblock Spherenet: Learning spherical representations for detection and classification in omnidirectional images.
\newblock In \emph{Proceedings of the European conference on computer vision (ECCV)}, pp.\  518--533, 2018.

\bibitem[Das et~al.(2023)Das, Goyal, Lee, Bhattacharjee, and Ganguly]{das2023crysmmnet}
Das, K., Goyal, P., Lee, S.-C., Bhattacharjee, S., and Ganguly, N.
\newblock Crysmmnet: multimodal representation for crystal property prediction.
\newblock In \emph{Uncertainty in Artificial Intelligence}, pp.\  507--517. PMLR, 2023.

\bibitem[Dorri et~al.(2018)Dorri, Kanhere, and Jurdak]{dorri2018multi}
Dorri, A., Kanhere, S.~S., and Jurdak, R.
\newblock Multi-agent systems: A survey.
\newblock \emph{Ieee Access}, 6:\penalty0 28573--28593, 2018.

\bibitem[Fey \& Lenssen(2019)Fey and Lenssen]{torch_geo}
Fey, M. and Lenssen, J.~E.
\newblock Fast graph representation learning with {PyTorch Geometric}.
\newblock In \emph{ICLR Workshop on Representation Learning on Graphs and Manifolds}, 2019.

\bibitem[Fuchs et~al.(2020)Fuchs, Worrall, Fischer, and Welling]{fuchs2020se}
Fuchs, F., Worrall, D., Fischer, V., and Welling, M.
\newblock Se (3)-transformers: 3d roto-translation equivariant attention networks.
\newblock \emph{Advances in neural information processing systems}, 33:\penalty0 1970--1981, 2020.

\bibitem[Gaus et~al.(2011)Gaus, Cui, and Elstner]{dft4}
Gaus, M., Cui, Q., and Elstner, M.
\newblock Dftb3: Extension of the self-consistent-charge density-functional tight-binding method (scc-dftb).
\newblock \emph{Journal of chemical theory and computation}, 7\penalty0 (4):\penalty0 931--948, 2011.

\bibitem[Goyal et~al.(2014)Goyal, Qian, Irle, Lu, Roston, Mori, Elstner, and Cui]{goyal2014molecular}
Goyal, P., Qian, H.-J., Irle, S., Lu, X., Roston, D., Mori, T., Elstner, M., and Cui, Q.
\newblock Molecular simulation of water and hydration effects in different environments: Challenges and developments for dftb based models.
\newblock \emph{The Journal of Physical Chemistry B}, 118\penalty0 (38):\penalty0 11007--11027, 2014.

\bibitem[Hern{\'a}ndez et~al.(1996)Hern{\'a}ndez, Gillan, and Goringe]{dft3}
Hern{\'a}ndez, E., Gillan, M., and Goringe, C.
\newblock Linear-scaling density-functional-theory technique: The density-matrix approach.
\newblock \emph{Physical Review B}, 53\penalty0 (11):\penalty0 7147, 1996.

\bibitem[Hourahine et~al.(2020)Hourahine, Aradi, Blum, Bonafe, Buccheri, Camacho, Cevallos, Deshaye, Dumitric{\u{a}}, Dominguez, et~al.]{hourahine2020dftb+}
Hourahine, B., Aradi, B., Blum, V., Bonafe, F., Buccheri, A., Camacho, C., Cevallos, C., Deshaye, M., Dumitric{\u{a}}, T., Dominguez, A., et~al.
\newblock Dftb+, a software package for efficient approximate density functional theory based atomistic simulations.
\newblock \emph{The Journal of Chemical Physics}, 152\penalty0 (12), 2020.

\bibitem[Jia et~al.(2025)Jia, Zhou, Liu, Wang, Wang, Zhang, Liu, Wang, Ye, Amezawa, et~al.]{jia2025closed}
Jia, X., Zhou, Z., Liu, F., Wang, T., Wang, Y., Zhang, D., Liu, H., Wang, Y., Ye, S., Amezawa, K., et~al.
\newblock Closed-loop framework for discovering stable and low-cost bifunctional metal oxide catalysts for efficient electrocatalytic water splitting in acid.
\newblock \emph{Journal of the American Chemical Society}, 2025.

\bibitem[Kim et~al.(2025)Kim, Chen, Cheng, Gindulyte, He, He, Li, Shoemaker, Thiessen, Yu, et~al.]{pubchem2025}
Kim, S., Chen, J., Cheng, T., Gindulyte, A., He, J., He, S., Li, Q., Shoemaker, B.~A., Thiessen, P.~A., Yu, B., et~al.
\newblock Pubchem 2025 update.
\newblock \emph{Nucleic Acids Research}, 53\penalty0 (D1):\penalty0 D1516--D1525, 2025.

\bibitem[Kingma \& Ba(2014)Kingma and Ba]{adam}
Kingma, D.~P. and Ba, J.
\newblock Adam: A method for stochastic optimization.
\newblock \emph{arXiv preprint arXiv:1412.6980}, 2014.

\bibitem[Kohn \& Sham(1965)Kohn and Sham]{dft1}
Kohn, W. and Sham, L.~J.
\newblock Self-consistent equations including exchange and correlation effects.
\newblock \emph{Physical review}, 140\penalty0 (4A):\penalty0 A1133, 1965.

\bibitem[Li et~al.(2024)Li, Wang, Zeng, Wu, and Yang]{li2024survey}
Li, X., Wang, S., Zeng, S., Wu, Y., and Yang, Y.
\newblock A survey on llm-based multi-agent systems: workflow, infrastructure, and challenges.
\newblock \emph{Vicinagearth}, 1\penalty0 (1):\penalty0 9, 2024.

\bibitem[Lin et~al.(2019)Lin, Lu, and Ying]{dft2}
Lin, L., Lu, J., and Ying, L.
\newblock Numerical methods for kohn--sham density functional theory.
\newblock \emph{Acta Numerica}, 28:\penalty0 405--539, 2019.

\bibitem[Luo et~al.(2024)Luo, Wang, Wang, Lv, Wang, Wang, and Ma]{luo2024crystalflow}
Luo, X., Wang, Z., Wang, Q., Lv, J., Wang, L., Wang, Y., and Ma, Y.
\newblock Crystalflow: A flow-based generative model for crystalline materials.
\newblock \emph{arXiv preprint arXiv:2412.11693}, 2024.

\bibitem[Michel et~al.(2018)Michel, Ferber, and Drogoul]{michel2018multi}
Michel, F., Ferber, J., and Drogoul, A.
\newblock Multi-agent systems and simulation: A survey from the agent commu-nity’s perspective.
\newblock In \emph{Multi-Agent Systems}, pp.\  17--66. CRC Press, 2018.

\bibitem[Orio et~al.(2009)Orio, Pantazis, and Neese]{orio2009density}
Orio, M., Pantazis, D.~A., and Neese, F.
\newblock Density functional theory.
\newblock \emph{Photosynthesis Research}, 102:\penalty0 443--453, 2009.

\bibitem[Oviedo et~al.(2022)Oviedo, Ferres, Buonassisi, and Butler]{oviedo2022interpretable}
Oviedo, F., Ferres, J.~L., Buonassisi, T., and Butler, K.~T.
\newblock Interpretable and explainable machine learning for materials science and chemistry.
\newblock \emph{Accounts of Materials Research}, 3\penalty0 (6):\penalty0 597--607, 2022.

\bibitem[Paszke et~al.(2017)Paszke, Gross, Chintala, Chanan, Yang, DeVito, Lin, Desmaison, Antiga, and Lerer]{pytorch}
Paszke, A., Gross, S., Chintala, S., Chanan, G., Yang, E., DeVito, Z., Lin, Z., Desmaison, A., Antiga, L., and Lerer, A.
\newblock Automatic differentiation in pytorch.
\newblock In \emph{Advances in Neural Information Processing Systems Workshop on Autodiff}, 2017.

\bibitem[Polat et~al.(2024)Polat, Kurban, and Kurban]{polat2024multimodal}
Polat, C., Kurban, M., and Kurban, H.
\newblock Multimodal neural network-based predictive modeling of nanoparticle properties from pure compounds.
\newblock \emph{Machine Learning: Science and Technology}, 5\penalty0 (4):\penalty0 045062, 2024.

\bibitem[Polat et~al.(2025{\natexlab{a}})Polat, Kurban, and Kurban]{polat2025quantumshellnet}
Polat, C., Kurban, H., and Kurban, M.
\newblock Quantumshellnet: ground-state eigenvalue prediction of materials using electronic shell structures and fermionic properties via convolutions.
\newblock \emph{Computational Materials Science}, 246:\penalty0 113366, 2025{\natexlab{a}}.

\bibitem[Polat et~al.(2025{\natexlab{b}})Polat, Kurban, Serpedin, and Kurban]{polat2025understanding}
Polat, C., Kurban, H., Serpedin, E., and Kurban, M.
\newblock Understanding the capabilities of molecular graph neural networks in materials science through multimodal learning and physical context encoding, 2025{\natexlab{b}}.
\newblock URL \url{https://arxiv.org/abs/2505.12137}.

\bibitem[Radford et~al.(2021)Radford, Kim, Hallacy, Ramesh, Goh, Agarwal, Sastry, Askell, Mishkin, Clark, et~al.]{clip}
Radford, A., Kim, J.~W., Hallacy, C., Ramesh, A., Goh, G., Agarwal, S., Sastry, G., Askell, A., Mishkin, P., Clark, J., et~al.
\newblock Learning transferable visual models from natural language supervision.
\newblock In \emph{International conference on machine learning}, pp.\  8748--8763. PmLR, 2021.

\bibitem[Ramakrishnan et~al.(2014)Ramakrishnan, Dral, Rupp, and von Lilienfeld]{ramakrishnan2014quantum}
Ramakrishnan, R., Dral, P.~O., Rupp, M., and von Lilienfeld, O.~A.
\newblock Quantum chemistry structures and properties of 134 kilo molecules.
\newblock \emph{Scientific Data}, 1, 2014.

\bibitem[Reiser et~al.(2022)Reiser, Neubert, Eberhard, Torresi, Zhou, Shao, Metni, van Hoesel, Schopmans, Sommer, et~al.]{reiser2022graph}
Reiser, P., Neubert, M., Eberhard, A., Torresi, L., Zhou, C., Shao, C., Metni, H., van Hoesel, C., Schopmans, H., Sommer, T., et~al.
\newblock Graph neural networks for materials science and chemistry.
\newblock \emph{Communications Materials}, 3\penalty0 (1):\penalty0 93, 2022.

\bibitem[Ruddigkeit et~al.(2012)Ruddigkeit, Van~Deursen, Blum, and Reymond]{ruddigkeit2012enumeration}
Ruddigkeit, L., Van~Deursen, R., Blum, L.~C., and Reymond, J.-L.
\newblock Enumeration of 166 billion organic small molecules in the chemical universe database gdb-17.
\newblock \emph{Journal of chemical information and modeling}, 52\penalty0 (11):\penalty0 2864--2875, 2012.

\bibitem[Satorras et~al.(2021)Satorras, Hoogeboom, and Welling]{satorras2021n}
Satorras, V.~G., Hoogeboom, E., and Welling, M.
\newblock E (n) equivariant graph neural networks.
\newblock In \emph{International conference on machine learning}, pp.\  9323--9332. PMLR, 2021.

\bibitem[Sch{\"u}tt et~al.(2021)Sch{\"u}tt, Unke, and Gastegger]{painn}
Sch{\"u}tt, K., Unke, O., and Gastegger, M.
\newblock Equivariant message passing for the prediction of tensorial properties and molecular spectra.
\newblock In \emph{International Conference on Machine Learning}, pp.\  9377--9388. PMLR, 2021.

\bibitem[Song et~al.(2025)Song, Luo, Zhang, Chen, Huang, Cao, Zhu, Liu, Zhang, Zou, et~al.]{song2025multiagent}
Song, T., Luo, M., Zhang, X., Chen, L., Huang, Y., Cao, J., Zhu, Q., Liu, D., Zhang, B., Zou, G., et~al.
\newblock A multiagent-driven robotic ai chemist enabling autonomous chemical research on demand.
\newblock \emph{Journal of the American Chemical Society}, 147\penalty0 (15):\penalty0 12534--12545, 2025.

\bibitem[Th{\"o}lke \& De~Fabritiis(2022)Th{\"o}lke and De~Fabritiis]{tholke2022torchmd}
Th{\"o}lke, P. and De~Fabritiis, G.
\newblock Torchmd-net: equivariant transformers for neural network based molecular potentials.
\newblock \emph{arXiv preprint arXiv:2202.02541}, 2022.

\bibitem[Van~der Hoek \& Wooldridge(2008)Van~der Hoek and Wooldridge]{van2008multi}
Van~der Hoek, W. and Wooldridge, M.
\newblock Multi-agent systems.
\newblock \emph{Foundations of Artificial Intelligence}, 3:\penalty0 887--928, 2008.

\bibitem[Wolf et~al.(2019)Wolf, Debut, Sanh, Chaumond, Delangue, Moi, Cistac, Rault, Louf, Funtowicz, et~al.]{wolf2019huggingface}
Wolf, T., Debut, L., Sanh, V., Chaumond, J., Delangue, C., Moi, A., Cistac, P., Rault, T., Louf, R., Funtowicz, M., et~al.
\newblock Huggingface's transformers: State-of-the-art natural language processing.
\newblock \emph{arXiv preprint arXiv:1910.03771}, 2019.

\bibitem[Xie \& Liu(2017)Xie and Liu]{xie2017multi}
Xie, J. and Liu, C.-C.
\newblock Multi-agent systems and their applications.
\newblock \emph{Journal of International Council on Electrical Engineering}, 7\penalty0 (1):\penalty0 188--197, 2017.

\bibitem[Xie et~al.(2021)Xie, Fu, Ganea, Barzilay, and Jaakkola]{xie2021crystal}
Xie, T., Fu, X., Ganea, O.-E., Barzilay, R., and Jaakkola, T.
\newblock Crystal diffusion variational autoencoder for periodic material generation.
\newblock \emph{arXiv preprint arXiv:2110.06197}, 2021.

\bibitem[Zhang et~al.(2025)Zhang, Wang, Zhu, Liu, Lin, and Cambria]{zhang2025mars}
Zhang, J., Wang, Z., Zhu, H., Liu, J., Lin, Q., and Cambria, E.
\newblock Mars: A multi-agent framework incorporating socratic guidance for automated prompt optimization.
\newblock \emph{arXiv preprint arXiv:2503.16874}, 2025.

\end{thebibliography}
\bibliographystyle{icml2025}




\end{document}